 \documentclass[final,5p,times,twocolumn]{elsarticle}

\usepackage{epsfig} 
\usepackage{graphics}
\usepackage{xspace}
\usepackage{amsmath}
\usepackage{subfigure}
\usepackage{booktabs}
\usepackage{color}

\definecolor{Red}{rgb}{0.7,0.0,0.0}
\definecolor{Green}{rgb}{0.0,0.7,0.0}
\definecolor{Blue}{rgb}{0.0,0.0,0.7}

\usepackage{amssymb}
\journal{Physics Letters A}

\begin{document}

\begin{frontmatter}



\title{On the agreement between small-world-like OFC model and real earthquakes}


\author[ifrj,on]{Douglas S. R. Ferreira\corref{cor}}
\ead{douglas.ferreira@ifrj.edu.br}

\author[on,uerj]{Andr\'es R. R. Papa\corref{cor}}
\ead{papa@on.br}

\author[fit]{Ronaldo Menezes}
\ead{rmenezes@cs.fit.edu}

\cortext[cor]{Corresponding authors}

\address[ifrj]{Instituto Federal de Educa\c{c}\~ao, Ci\^encia e Tecnologia do Rio de Janeiro, Paracambi, RJ, Brazil}

\address[on]{Geophysics Department, Observat\'orio Nacional, Rio de Janeiro, RJ, Brazil}

\address[uerj]{Instituto de F\'isica, Universidade do Estado do Rio de Janeiro, Rio de Janeiro, RJ, Brazil}

\address[fit]{BioComplex Laboratory, Computer Sciences, Florida Institute
  of Technology, Melbourne, USA}

\begin{abstract}
In this article we implemented simulations of the OFC model for earthquakes for two different topologies: regular and small-world, where in the latter the links are randomly rewired with probability $p$ . In both topologies, we have studied the distribution of time intervals between consecutive earthquakes and the border effects present in each one. In addition, we also have characterized the influence that the probability $p$ produces in certain characteristics of the lattice and in the intensity of border effects. From the two topologies, networks of consecutive epicenters were constructed, that allowed us to analyze the distribution of connectivities of each one. In our results distributions arise  belonging to a family of non-traditional distributions functions, which agrees with previous studies using data from actual earthquakes. Our results reinforce the idea that the Earth is in a critical self-organized state and furthermore point towards temporal and spatial correlations between earthquakes in different places.
\end{abstract}


\end{frontmatter}





\section{Introduction}
\label{intro}
%Uma das formas largamente utilizadas pela física estatística em estudos sismológicos é a análise de propriedades estatísticas de terremotos. Dois resultados empíricos amplamente conhecidos, obtidos através destas análises, são: a lei de Gutenberg-Richter para a distribuição de tamanhos de terremotos~\cite{gutenberg1956earthquake}, e a lei de Omori para a evolução temporal da frequência de aftershocks~\cite{omori1894aftershocks}. Uma propriedade importante observada nestas duas distribuições --- e também em outras distribuições relacionadas aos sismos --- é que estas são caracterizadas por leis de potência, o que abre espaço para a interpretação de que a crosta terrestre se encontra em um estado crítico, e se comporta como um sistema complexo auto-organizado.

The concept of self-organized criticality (SOC), originally presented by P. Bak, C. Tang and K. Wiesenfeld~\cite{bak1987self} and widely used in statistical physics refers, generally, to the property that a large class of dynamical systems has to organize spontaneously into a dynamic critical state without the need for any fine tuning of some external control parameter. A signature of self-organized criticality in a system is the invariance of temporal and spatial scales, observed by power-law distributions and finite size scaling.

In view of the above characteristics, the concept of SOC has been employed in various fields of science such as neurobiology~\cite{linkenkaer2001long}, economics~\cite{mantegna1995scaling}, plasma physics~\cite{nurujjaman2007realization} and geophysics~\cite{olami1992self}, among many others.

In the field of geophysics, one of the methods most used by statistical physics in seismological studies is the analysis of statistical properties of earthquakes. Two widely known empirical results obtained through these analyzes are: the Gutenberg-Richter law for the distribution of earthquakes' sizes~\cite{gutenberg1956earthquake}, and the Omori law for the time evolution of the frequency of aftershocks~\cite{omori1894aftershocks}. An important property of these two observed distributions, and also other distributions with respect to earthquakes, is that they are characterized by power laws. Thus, the assumption that the Earth's crust is in a critical state and behaves like a self-organized complex system emerges as a possible explanation for the occurrence of the long-range spatio-temporal links present in earthquakes dynamics, given that in a self-organized system the occurrence of partial synchronization of the elements leads to the creation of long-range relationships.

Despite all the existing knowledge about the production of seismic waves through slips on faults, much remains to be discovered regarding the dynamics responsible for these slips. A key step in deepening this knowledge is the study, analysis and modeling of the seismic distributions in space and time. Following this line of reasoning, several models that incorporate characteristics of SOC have been used to reproduce properties found in earthquakes, among which stands out the model developed by Olami, Feder and Christensen (known as OFC model), which has played an important role in the phenomenological study of earthquakes, because, despite being one of the simplest models in the universe of self-organized critical models (even though there is some controversy regarding its self-organized criticality~\cite{de2000self,lise2001self}), it displays a phenomenology similar to the one found in actual earthquakes, with satisfactory results in reproducing distributions as the Gutenberg-Richter law and the Omori law~\cite{christensen1992scaling,christensen1992variation}.

The OFC model with a regular topology and \textit{open} boundary conditions presents an inhomogeneity in relation with the border of the lattice, coming from the fact that the biggest avalanches in the system are caused from sites located near the borders~\cite{grassberger1994efficient,lise2001scaling,peixoto2006network}, which will cause the number of times a site becomes the epicenter to vary based on how close this site is in relation to the border of the lattice, occurring more frequently in sites located closer to the borders. This \textit{border effect}, as named previously~\cite{peixoto2006network}, will influence directly in the analysis of the OFC model, therefore, it will be important to know how this effect extends through the lattice.

In recent years several studies have pointed to a possible long-range spatial and temporal correlation between earthquakes~\cite{crescentini1999constraints,parsons2002global,mega2003power,varotsos2005similarity,ferreira2014smallworld}. So, we can understand that when a seismic event occurs, this event may induce a redistribution of tensions through the Earth's crust in such a way that would cause other seismic events not only in times and places nearby, but also in distant times and places. In order to introduce this line of reasoning in the OFC model, it was proposed in~\cite{caruso2006olami} a different topology for the lattice, called \textit{small-world topology}, which has a fraction of the links randomly rewired between different sites in the lattice, simulating a kind of ``small-world effect''~\cite{watts1998collective}.

An alternative and powerful tool in the study of earthquakes is the so-called network of successive epicenters which was originally introduced by Abe and Suzuki in real local seismic data~\cite{abe2004scalefree}, subsequently used by Peixoto and Prado in synthetic data obtained by using the OFC model in its original form~\cite{peixoto2004distribution} and very recently by Ferreira \textit{et al.}~\cite{ferreira2014smallworld} for the whole Earth earthquake catalog. 

Aiming to contribute to the understanding of earthquake dynamics, in this paper we first carried out a study on the distribution of time intervals between consecutive earthquakes under the views of two different lattice topologies: regular (which has a fashion similar to the original OFC) and small-world. After this, we have built a complex network of successive epicenters from synthetic catalogs produced with the OFC model, using both regular and small-world topologies. Thus we were able to study the characteristics of the distribution of connectivities of each of the created epicenters networks. We have also studied ``border effects'' present in each of the topologies, as well as the influences and consequences produced by the small-world topology. Our results were also compared with results from previous studies of our and others groups using real and synthetic data.

\section{The OFC model, topology and boundaries}
\label{ofc_model}
Introduced in 1992 by Olami, Feder and Christensen, the OFC model proposes a cellular automata model based on a simplification of the spring- block model of Burridge and Knopoff (BK model)~\cite{burridge1967model}. The OFC model can be represented by a bi-dimensional square network of blocks interconnected by springs, where each block is also connected through a spring to a single rigid driven plate. The blocks are also connected by friction to other rigid fixed plate on which they stay. Due to the relative motion between the plates (imposed in the model), all the blocks will be subjected to an elastic force which tends to put them in motion and other frictional force opposite to the first. When the resulting force in one of the blocks is greater than the maximum static friction force, the block slides and relaxes to a position of zero force, so that there is a rearrangement of forces in its first neighbors, which can cause other slippages and the emergence of a chain reaction.

Such a model is simulated considering a square $L \times L$ lattice with $N = L^2$ sites, where to each site with coordinates $(i,j)$ is assigned a force value $F_{i,j}$. Initially, the value for each site is randomly chosen between $0$ and $F_{th}$, where $F_{th}$ is the limit value for the friction force (maximum static friction). The values of the forces at sites are then increased at a constant rate simultaneously and uniformly throughout the lattice, i.e., all sites increase the values of $F_{i,j}$ with the same speed, $dF_{i,j}/dt = v$, which can be regarded as a simulation of a uniform tectonic loading. When a site $(i,j)$ reaches a tension value equal to or greater than $F_{th}$, thus becoming unstable, the force $F_{i,j}$ is redistributed among its nearest neighbors and after that taken equal to zero, as can be seen in the relaxation rule governing when $F_{i,j} \geq F_{th}$:
%
%\begin{equation}
%\label{S}
%F_{i,j} \geq F_{th} \Rightarrow \left\{
%\begin{array}{ll}   % they are two tiny letters "L"  inside the keys, to indicate "left-justified"
%F_{n,n} \rightarrow F_{n,n} + \alpha F_{i,j}
%\\
%F_{i,j} \rightarrow 0
%\end{array} \right.		%do not forget to insert the dot "." after the command "\right"
%\end{equation}
%
\begin{equation}
\label{S}
\begin{array}{ll}   % they are two tiny letters "L"  inside the keys, to indicate "left-justified"
F_{n,n} \rightarrow F_{n,n} + \alpha F_{i,j} \, \,,
\\
F_{i,j} \rightarrow 0
\end{array} 
\end{equation}
where $F_{n,n}$ are the forces on the set ($n,n$) of first neighbors of the site $(i,j)$ and $\alpha$ controls the dissipation intensity, where $\alpha = 1/4$ corresponds to the conservative case and $0 < \alpha < 1/4$ to dissipative cases. In this work we study \textit{open} boundary conditions, in which all blocks of the lattice have the same $\alpha$ value. If the redistribution of forces causes some other site reaches a value equal to or greater than the threshold $F_{th}$, the process of sliding and relaxation will be repeated until all sites are below this limit. When this occurs we say that the earthquake has evolved completely. The energy released by the earthquake (also called magnitude in some works), measured by its size $s$, will be given by the total number of block slides, from the initial slip.

After the complete evolution of an earthquake, the plates continue to move one relatively to the other until another event occurs. So, to find the site in the system where the next earthquake will start, we look for the site with the larger $F$ value ($F_{larg}$) and add the difference $F_{th} - F_{larg}$ to the values of $F$ in all sites, which will make the process of redistribution and relaxation begins again. It is important to emphasize that because of the time interval between two seismic events be much longer than the duration of an event, earthquakes will be considered as instantaneous events, which means that during the relaxation of the site and consequent redistribution of forces, it is considered that the uniform driving stops, or in other words, that no extra time is added during this process.
%
%simulando assim a continuidade do movimento relativo entre as placas até que novamente algum dos blocos fique sujeito a uma força maior do que o atrito estático.
%

\begin{figure}[t]
  \begin{center}
    \subfigure[] {
    \includegraphics[width=0.47\columnwidth]{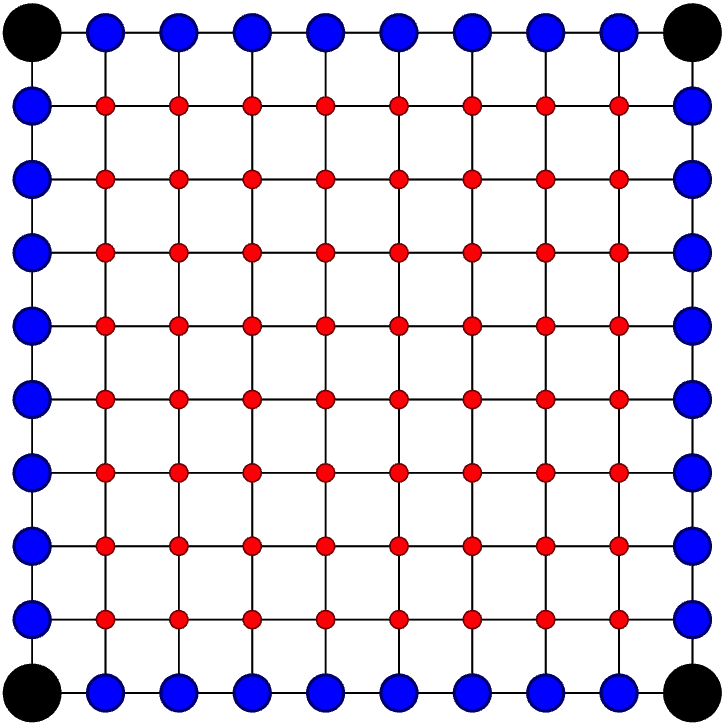}
      \label{net_regular}}
    \subfigure[]  {
    \includegraphics[width=0.47\columnwidth]{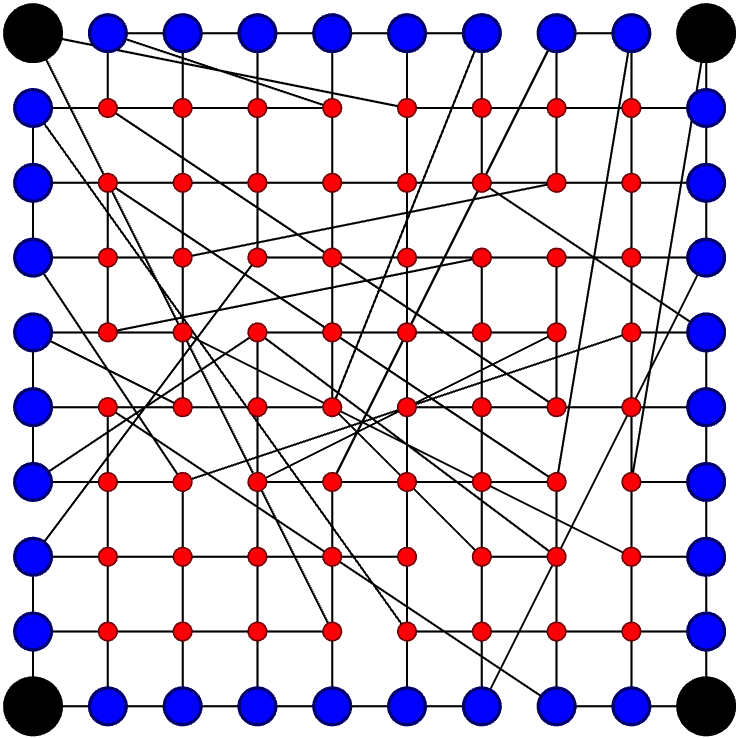}
      \label{net_smallworld}}
    \caption{Example of a $10 \times 10$ lattice with \textit{open} boundary conditions, where each site in black (small size), blue (medium size) and red (large size) has 2, 3 and 4 neighbors, respectively. \subref{net_regular} Regular topology. \subref{net_smallworld} Small-world topology with $p=0.05$.}
    \label{networks}
  \end{center}
\end{figure}

Based on the method of Watts and Strogatz to build networks~\cite{watts1998collective}, a new ``small-world topology'' can be used for the lattice in the OFC model instead of the regular topology. This new small-world topology has its construction explained below. Starting from a regular two-dimensional lattice, where each site is connected to its nearest neighbors, each edge of the lattice is randomly rewired with probability $p$, keeping fixed the original connectivity of each site. This means that for the link $i_1 - i_2$, which connects the site $i_1$ to one of its nearest neighbors $i_2$, there is a probability $p$ for this link to be rewired. If the link is chosen, another link $j_1 - j_2$, that connects the site $j_1$ to one of its nearest neighbors $j_2$, will be randomly selected, and the links $i_1 - i_2$ and $j_1 - j_2$ will be replaced by the pairs $i_1 - j_2$ and $j_1 - i_2$, respectively. The procedure is then repeated for all the links and the result is a lattice that is between a regular grid ($p=0$) and a random network ($p=1$), as shown in the example of Figure~\ref{networks} for $p=0.05$. Therefore, it can be seen that with this procedure, it is possible to create connections between very far sites on a lattice with regular topology, thus simulating some type of long-range correlation.

To construct a network of epicenters, we must first define what will be called epicenter in our model. Analogously to what was done in~\cite{peixoto2004distribution}, we define as epicenter each site that starts a new avalanche, regardless of the size of this avalanche. Thus, each epicenter defines a network vertex and the creation of edges will follow the temporal order of the epicenters, i.e., each site-epicenter will be connected to the following site-epicenter, thus forming a directed network of successive epicenters, where the degree of connectivity $k$ of each vertex is given by the number of edges that ``arrive'' and ``leave'' each vertex. It should be noted that if two subsequent events occur at the same site, this vertex is connected to itself by a loop or self-edge. The occurrence of repetitions of the same sequence of epicenters is also possible, therefore, parallel vertices are allowed. Another feature of this epicenter network is that due to its formation process, for every in-edge in a given vertex there will be an out-edge in the same vertex, which will cause the degree of connectivity values of incoming and output at each site to always be  the same, making the overall degree of connectivity $k$ of all vertices of the network to always be an even number (the only exceptions are the first and last vertices of the sequence of epicenters).

\section{Results and Discussion}
\label{results}
As previously mentioned, in this study we compare results obtained from the OFC model using regular topologies with the results for small-world topologies. These results are also compared to results obtained in previous studies using actual earthquake catalogs.

%as características obtidas nas redes de epicentros construídas a partir do modelo OFC com regular lattice e small-world lattice.

%\green{VER SE DE REPENTE É MELHOR COLOCAR ESTE SUB-ITEM DEPOIS DO Border effect, PARA EMENDAR NA PARTE DO TEXTO QUE DIZ ``mesmo uma pequena quantidade de ``rewired'' das conexões já é capaz de produzir mudanças significativas na dinâmica do sistema.'' COM ISSO JA COLOCARIA TAMBEM NESTA SECCAO UM GRAFICO DE VARIACAO DO TAMANHO DO EFEITO DE BORDA COM O VALOR DE $p$.}

As previously stated, the OFC model assumes that a constant variation of strength occurs in all sites of the system. Also, each event is considered as instantaneous, i.e., the time interval between two consecutive events is considered to be much greater than the duration of a single event. Therefore, we can assume that the amount $F_{th} - F_{larg}$ added to all lattice sites after the complete evolution of each earthquake is proportional to the value of the time interval between two consecutive earthquakes, which allows us to calculate the distribution of time intervals between consecutive events in our model. We will, therefore, observe the behavior of this distribution for the case in which the OFC model is implemented on the regular and small-world topologies. As stated in~\cite{caruso2006olami}, in the small-world case, probabilities above the value $p \simeq 0.1$ cause a progressive  disappearance in the power laws of the earthquakes size distributions, so we will adopt the same value for the rewire probability that was used in that work, i.e., $p=0.006$.

\begin{figure}[t]
  \begin{center}
    \subfigure[] {
    \includegraphics[width=0.75\columnwidth]{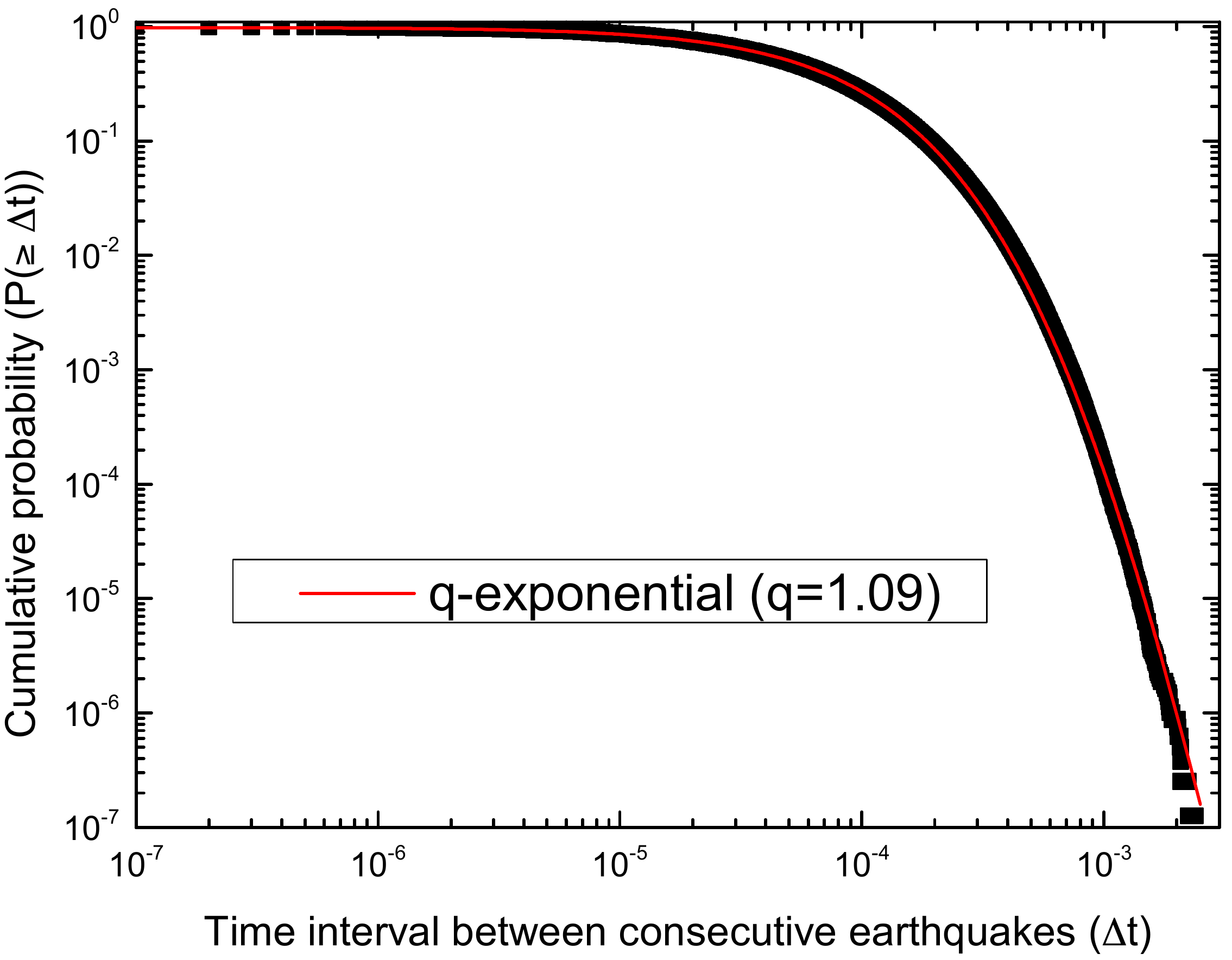}
      \label{D_smallworld}}
    \subfigure[]  {
    \includegraphics[width=0.75\columnwidth]{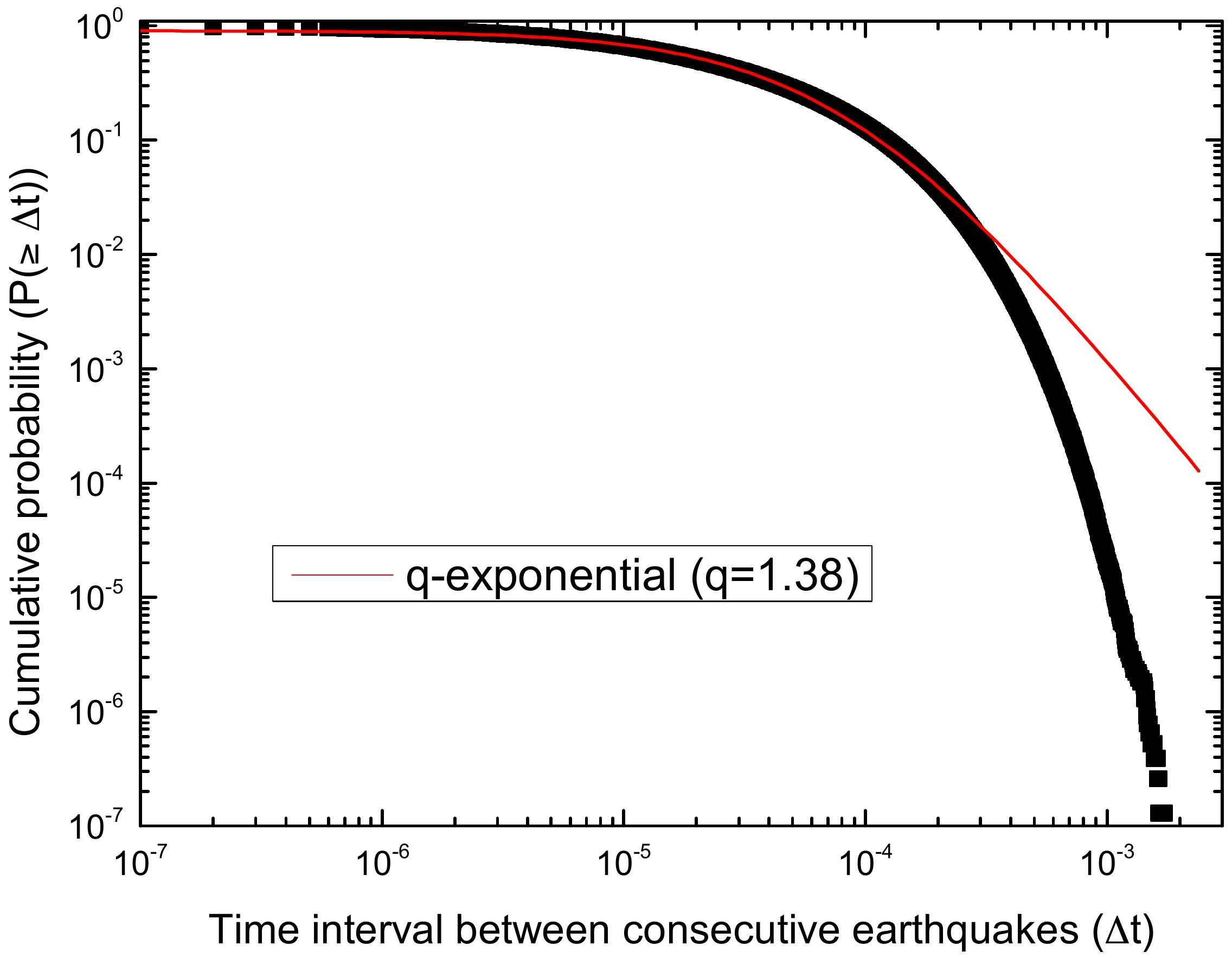}
      \label{D_regular}}
    \caption{Cumulative probability distribution of time intervals between consecutive earthquakes using lattices of size $L=200$ and $\alpha = 0.20$. Solid red lines represent adjustments to $q$-exponentials functions, where the best fittings were obtained for \subref{D_smallworld} $q=1.09  \pm 0.01$ and $\beta = 13\,711.4  \pm 3.0$ in small-world case and \subref{D_regular} $q=1.38 \pm 0.01$ and $\beta = 30\,164.6 \pm 29.2$ in the regular case. In both cases were considered $8 \times 10^{6}$ events after the transient regime and the time units are considered arbitrary.}
    \label{D}
  \end{center}
\end{figure}

Figure \ref{D_smallworld} presents the cumulative distribution of time intervals between successive events for the small-world topology. The best fit to this distribution is obtained by a non-traditional function of \textit{$q$-exponential} type, or $P(\geq \Delta t) = e^{-\beta \Delta t}_{q}$, with the \textit{$q$-exponential} function defined by:
\begin{equation}
\label{q_exp}
e_{q}^{x} = \left\{ 
  \begin{array}{l l}
    [1 + (1 - q)x]^{1/(1 - q)}	&  \text{if} \quad [1 + (1 - q)x] \geq 0\\
    0			 &  \text{if} \quad [1 + (1 - q)x] < 0
  \end{array} \right.
\end{equation}
where the limit $q \rightarrow 1$ recovers the standard exponential function. An important property is that if $q>1$, when $\Delta t \gg [- \beta (1-q)]^{-1}$, the cumulative probability distribution of inter-event intervals can be approximated to a power law given by $P(\geq \Delta t) \sim \Delta t^{1/(1-q)}$. It is noteworthy that the $q$-exponential distribution belongs to the family of ``Tsallis distributions'' and arises naturally from the maximization of Tsallis entropy, which is used to explain a variety of complex systems where the statistical mechanics of Boltzmann-Gibbs does not seem to apply~\cite{tsallis1988possible}. These systems have features such as, long-range interaction between its elements and temporal memory of long-range, making the Tsallis entropy -- and consequently the ``Tsallis distributions'' -- useful in diverse areas such as economy~\cite{borland2002option,queiros2005emergence,ludescher2011universal}, biology~\cite{tamarit1998sensitivity}, geophysics~\cite{barbosa2013statistical,vallianatos2013evidence,balasis2011universality,darooneh2010nonextensive}, astrophysics~\cite{livadiotis2011first,esquivel2010tsallis}, high energy physics~\cite{aad2011charged,gurtu2010transverse}, among many others. We draw attention here to the fact that this result has a remarkable agreement with the results obtained in~\cite{ferreira2014smallworld}, where catalogs of worldwide actual earthquakes were used and yielded a cumulative distribution of intervals between successive events obeying a $q$-exponential with index $q=1.08$. Similar results were also obtained using the catalogs of earthquakes in California and Japan, where the values of the index $q$ were equal to $1.13$ and $1.08$, respectively~\cite{abe2005scale}. It is added to this the further fact that for both California and Japan $q$-exponential functions are also found in distributions for distances between successive earthquakes~\cite{abe2003law}.

The Figure \ref{D_regular} shows the cumulative probability distribution found when a regular topology is used with the OFC model. When fitting a $q$-exponential function to this distribution it can be observed that there is no good agreement with data, meaning that for the regular case the cumulative probability distribution of inter-event intervals cannot be considered as a $q$-exponential.

\begin{figure}[t]
\begin{center}
\includegraphics[width=0.5\columnwidth]{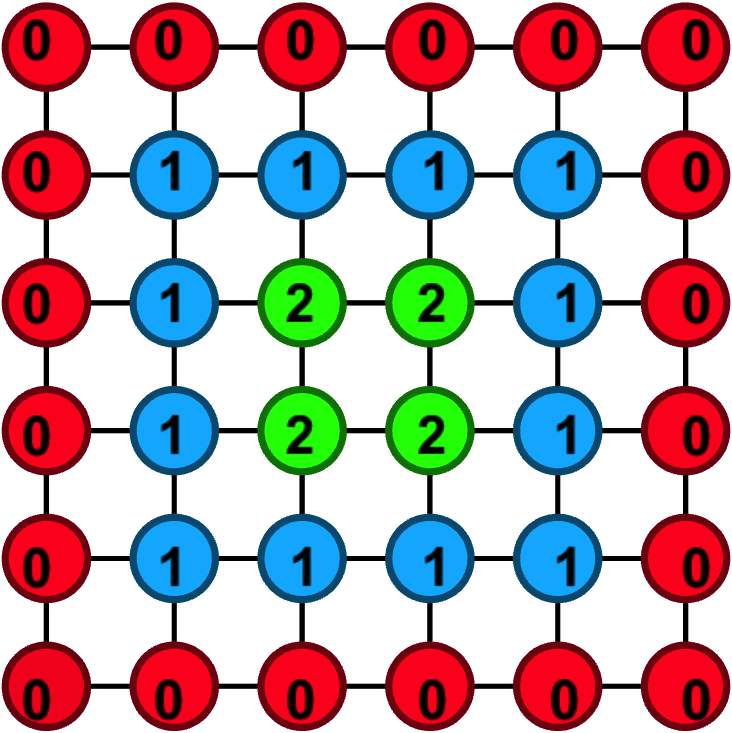}
\caption{Scheme of $6 \times 6$ lattice exemplifying three different layers. All sites belonging to ``Layer $0$'' (red) have distance to the border equal to $0$; sites belonging to ``Layer $1$'' (blue) have distance to the border equal to $1$; sites belonging to " Layer $2$" (green) have distance to the border equal $2$, and so on.}
\label{layer}
\end{center}
\end{figure}

We also have studied the spatial distribution of epicenters, i.e, how the epicenters are distributed throughout the lattice. However, before do that we introduce here the concept of ``layer'' as each set of sites in which all elements have the same distance to the nearest border, as shown in the scheme of Figure~\ref{layer}.

\begin{figure}[!t]
  \begin{center}
    \subfigure[] {
    \includegraphics[width=\columnwidth]{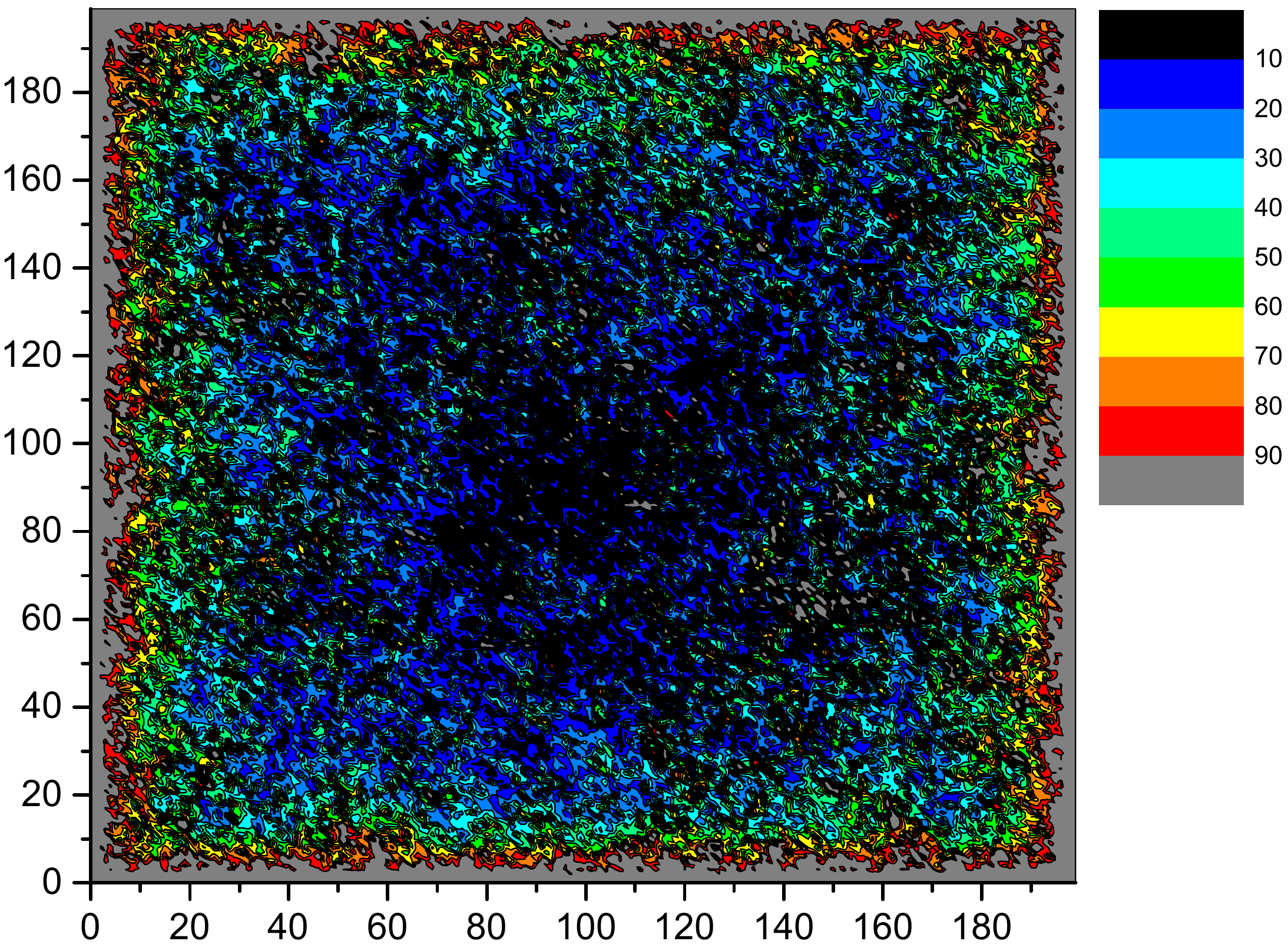}
      \label{map_regular}}
    \subfigure[]  {
    \includegraphics[width=\columnwidth]{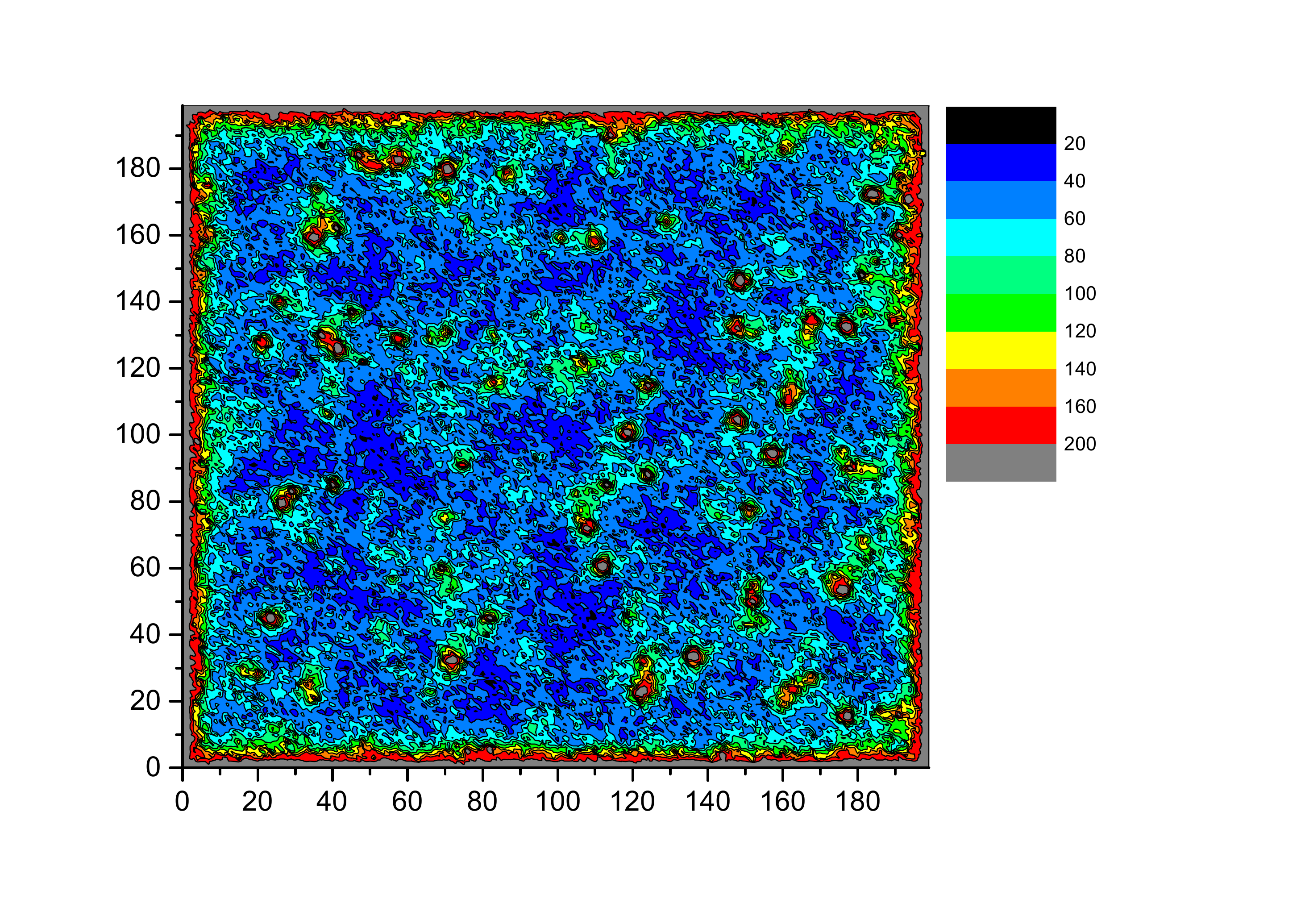}
      \label{map_smallworld}}
    \caption{Map of epicenters distribution through a lattice, i.e, the number of epicenters in each site of the lattice, for $L=200$ with conservation parameter $\alpha=0.20$. We have considered $8 \times 10^6$ events after the transient regime, but represented only earthquakes of size $s > 1$. \subref{map_regular} Regular lattice, where we can see clearly that the number of epicenters  decreases as it moves away from the borders. \subref{map_smallworld} Small-world lattice with $p=0.006$. The border effects are smaller, because after the layer number 20 (approximately) the distributions of epicenters is much more homogeneous.}
    \label{color_map}
  \end{center}
\end{figure}

In Figure~\ref{color_map} we have a distribution map of epicenters on a regular topology [Figure~\ref{map_regular}] and on a small-world topology [Figure~\ref{map_smallworld}]. It can be observed that the border effect is more intense in the regular case than in the small-world, or in other words, that this effect reaches a larger number of layers when using a regular topology than when using a small-world one. In order to obtain a quantity sufficient for statistical analysis we have considered $8 \times 10^{6}$ events after the transient regime. In addition, we have used only those epicenters of earthquakes with size $s >1$, due to the fact that events of size equal to $1$ seem to obey their own statistics~\cite{grassberger1994efficient}.
%tendo em vista que na rede small-world, para $p=0.006$, a frequência de epicentros se torna constante após uma distância de aproximadamente 20 sítios da borda, o que não ocorre na regular lattice. 

\begin{figure}[t]
\begin{center}
\includegraphics[width=\columnwidth]{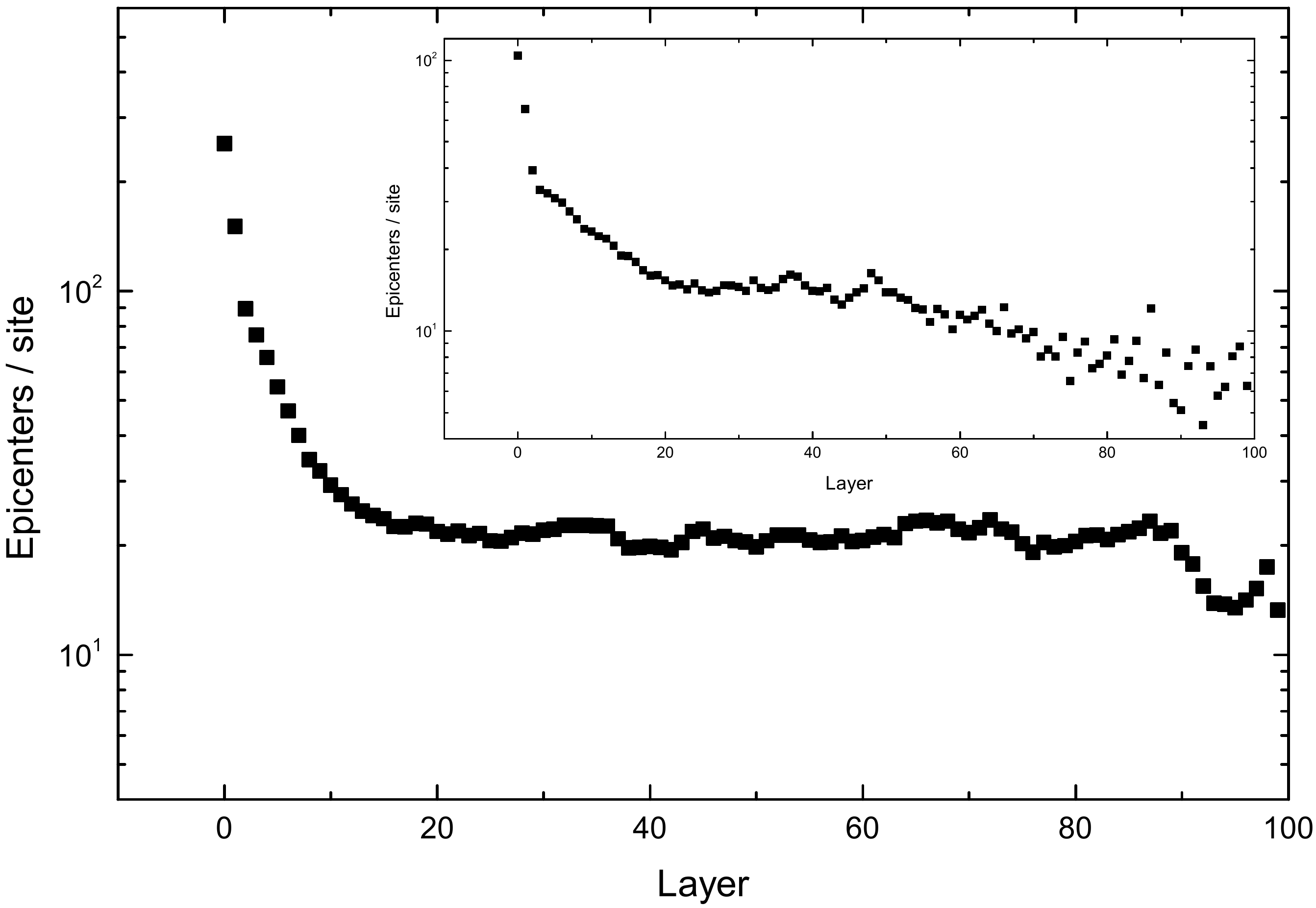}
\caption{Plot for the average number of epicenters occurring at each site by layer into a lattice with small-word topology of size $L=200$ with parameter $\alpha = 0.20$ and probability $p=0.006$. The abscissa axis represents the layer number and the ordinate axis represents the ratio between the number of epicenter occurring in this layer and the number of existing sites in the respective layer.  \textit{Inset}: represents the same graph but in a regular topology. In both cases we considered $8 \times 10^6$ events after the transient regime but used only earthquakes with $s > 1$.}
\label{mindist}
\end{center}
\end{figure}

The average quantitative difference in the intensity of border effects between regular and small-world topologies is shown in Figure~\ref{mindist}, which shows the distribution of the average number of epicenters occurring at each site of the lattice in each layer.
%
%Podemos então notar que no caso regular os efeitos se estendem por aproximadamente 100 camadas da lattice (resultado também encontrado anteriormente em~\cite{peixoto2006network}) enquanto que para o caso small-world estes efeitos se estendem por apenas 20 camadas aproximadamente. Convém salientar ainda que a probabilidade utilizada na construção da small-world lattice foi $p=0.006$, o que significa que mesmo uma pequena quantidade de ``rewired'' das conexões já é capaz de produzir mudanças significativas na dinâmica do sistema.
%
We can observe, from Figures \ref{color_map} and \ref{mindist}, that in the small-world case, border effects extend for less than 20 layers of the lattice, while for the regular case these effects remain relevant in a much larger fraction of the lattice (more specifically, the first 100 layers of the lattice, as reported in~\cite{peixoto2006network}). It is also noted that in these figures, the probability used in the construction of the lattice with small-word topology was $p=0.006$, meaning that even an extremely small amount of ``rewired'' connections is capable of producing significant changes in the system dynamics.

\begin{figure}[b]
\begin{center}
\includegraphics[width=0.7\columnwidth]{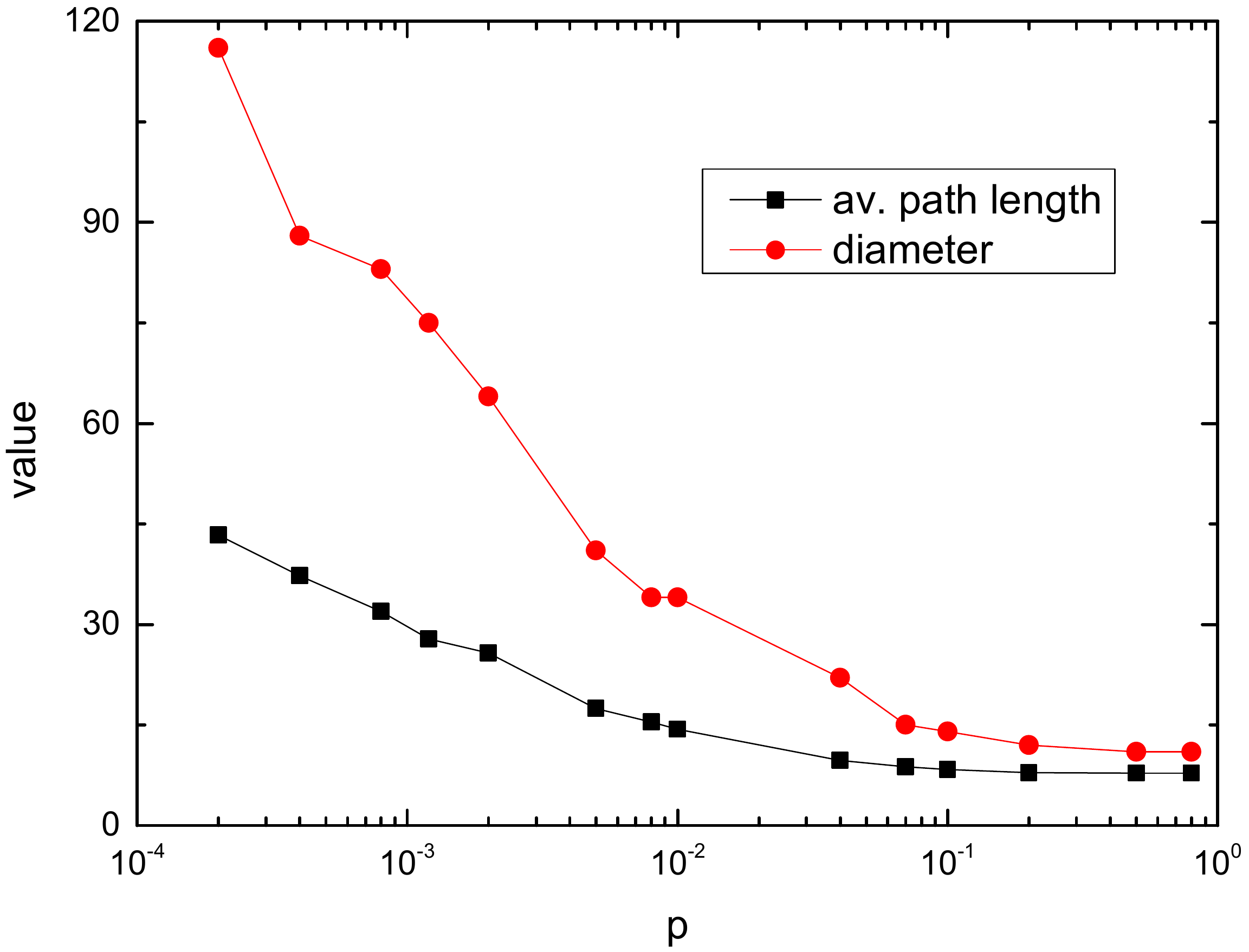}
\caption{\textit{Diameter} values (red circles) and \textit{average path length} (black squares) of a lattice with small-world topology due to the rewire probability $p$. We have used $L=100$ and conservative parameter $\alpha = 0.20$. Importantly, the result for the \textit{average path length} is already known~\cite{caruso2006olami}. }
\label{p_x_ell}
\end{center}
\end{figure}

To evaluate the influence of the parameter $p$ in the lattice we first measured the relationship of this parameter with the \textit{average path length} ($\ell$) of the lattice (similar to what was performed in~\cite{caruso2006olami}) and also with the \textit{diameter} ($d$) of the lattice (for further explanations on the \textit{average path length} and \textit{diameter} of a network see~\cite{albert2002statistical}). The results are shown in Figure~\ref{p_x_ell}, where it can be clearly seen that both $\ell$ and $d$ decrease significantly with the increasing of probability $p$. But from $p=0.1$ it is noted that the increase in the probability have virtually no influence on the \textit{average path length} and the \textit{diameter} of the lattice.

\begin{figure}[t]
\begin{center}
\includegraphics[width=0.7\columnwidth]{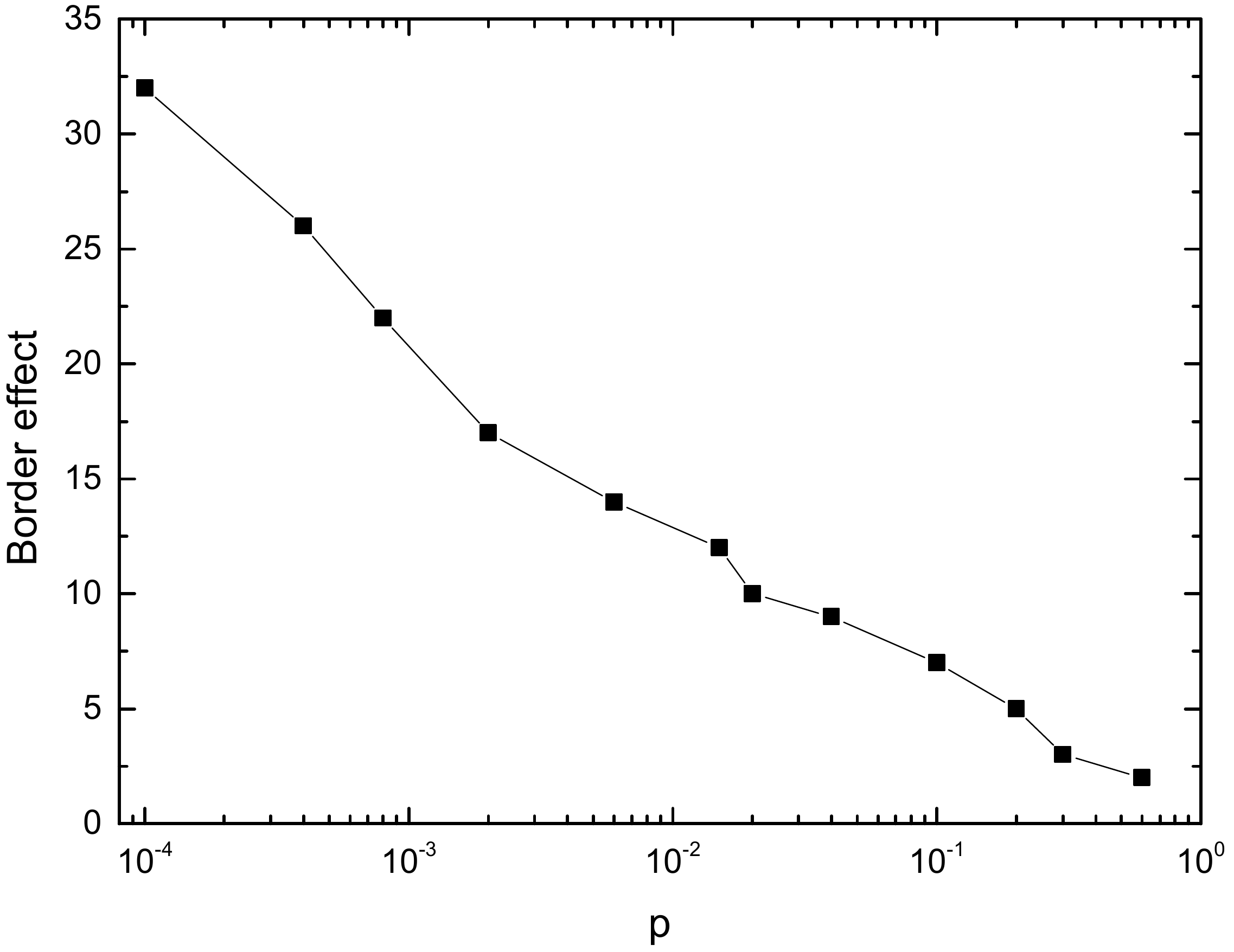}
\caption{Relationship between border effects and the rewire probability $p$ in a small-world topology for a lattice with size $L=100$ and $\alpha = 0.20$. The ordinate indicates the layer from which the border effect is no longer present, i.e.,  when the average number of epicenter at each site per layer becomes approximately constant. We considered $6 \times 10^6$ events after the transient regime.}
\label{p_x_bordereffect}
\end{center}
\end{figure}

Additionally, we investigated the way in which the rewire probability $p$ influences the border effect. From Figure \ref{p_x_bordereffect} can be noted that as the network becomes more ``small-world'', with the increase in the value of $p$, the border effect decreases.  It is possible to note that even very small probability values are able to produce significant reductions in the border effect, when compared to the regular case.

Before introducing our results regarding scale-free properties of the epicenters network, obtained from the model, it is important to highlight the result obtained in a previous study using real data from global catalog of earthquakes (available http://quake.geo.berkeley.edu/anss) covering earthquakes with magnitude greater than 4.5 (on the Richter scale) across the globe. In~\cite{ferreira2014towards} the authors created a network of epicenters using a time window model able to capture the small-world characteristics present in these epicenters networks. Thus, it was found that the distribution of connectivities for the sites-epicenters, has scale-free properties and obeys a function of the \textit{$q$-Gaussian} type, $f(x) = A[1 - (1 - q)\alpha x^2]^{1/(1 - q)} \equiv Ae^{-\alpha x^2}_{q}$ (which also belongs to the family of Tsallis distributions), where $e_{q}^{x}$ is the \textit{$q$-exponential} function. The $q$-Gaussian probability distribution is a generalization of the Gaussian curve, being dependent on the parameters $A$, $\alpha$ and the exponent $q$, where the limit $q \rightarrow 1$ recovers the standard Gaussian distribution. In the limit of large values of $x$ it is obtained again the power law approximation.

%\red{ALEM DISSO COLOCAR TAMBEM QUE NO LIMITE $x \gg [\alpha (1-q)]^{-1}$ (confirmar isso!!!!) SE $q>1$, TEMOS A APROXIMACAO PARA LEI DE POTENCIA $f(x) \sim x^{1/(1-q)}$}.

In our work, we use the model described earlier to generate catalogs of synthetic data and then build successive directed epicenters networks for the studied topologies: regular and small-world (with $p=0.006$), to obtain the connectivity distribution of vertices of each network. 

However, in order to keep the simulated system as small as possible, to minimize the border effects and to avoid also problems related to small size earthquakes, we have considered epicenters occurred in layers of order 20 or superior~\cite{lise2001scaling,peixoto2006network} and considered only those earthquakes with size 50 or greater.

\begin{figure}[t]
  \begin{center}
    \subfigure[] {
    \includegraphics[width=0.75\columnwidth]{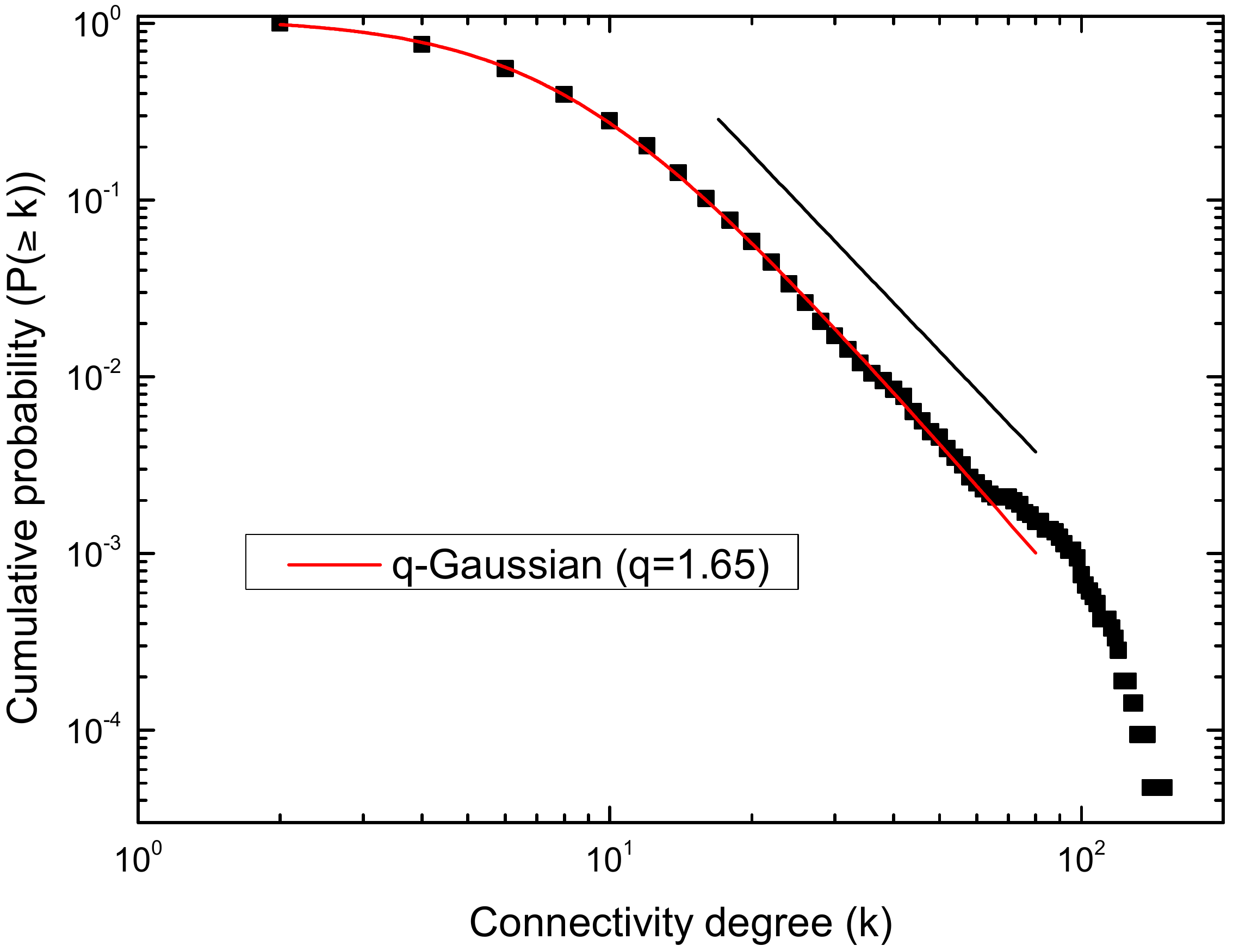}
      \label{k_smallworld}}
    \subfigure[]  {
    \includegraphics[width=0.75\columnwidth,]{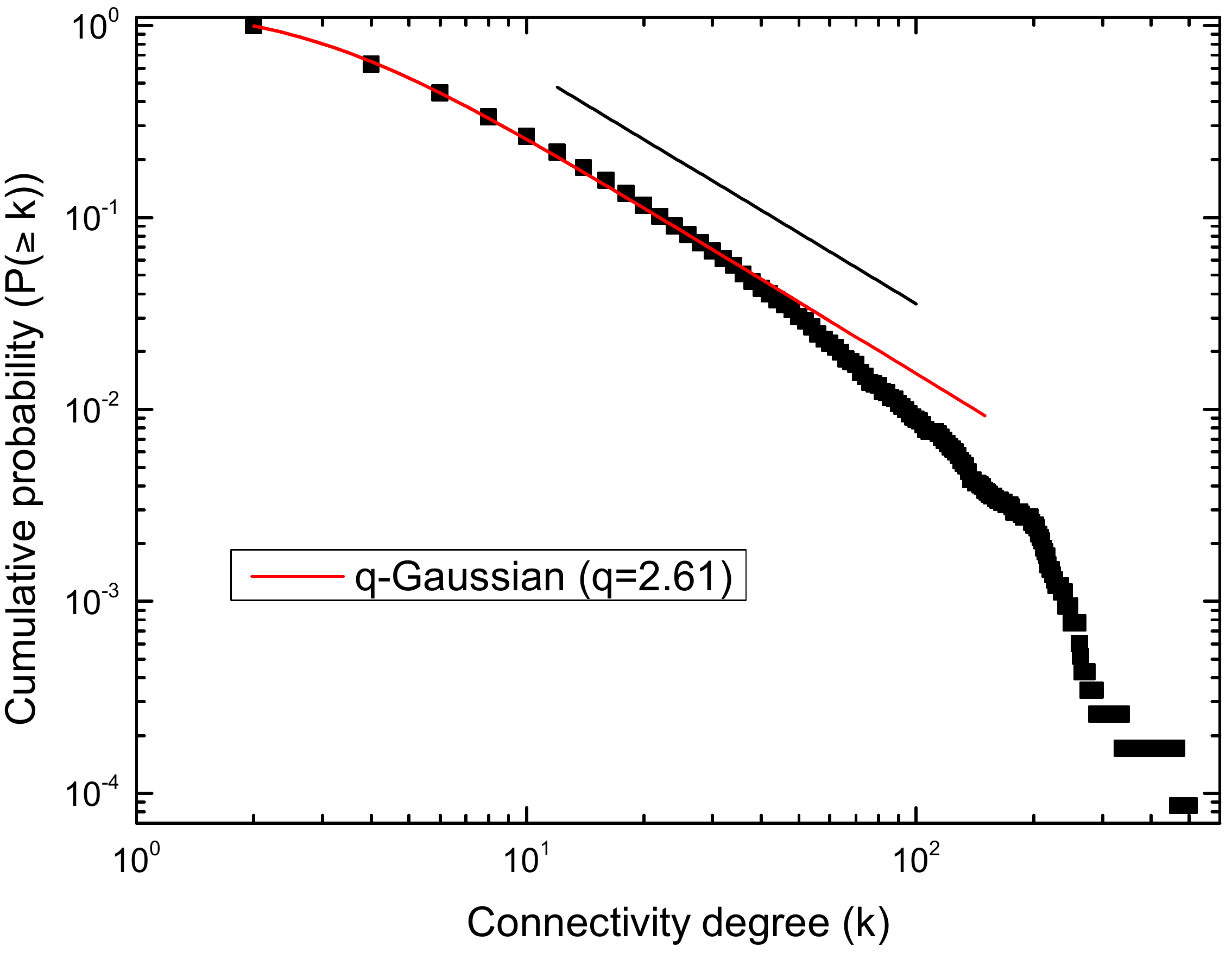}
      \label{k_regular}}
    \caption{Cumulative distribution of connectivities for lattices of size $L=200$ with $\alpha=0.20$. The statistics were made on $8 \times 10^6$ events after the transient regime, but were used only earthquakes of size $s \geq 50$. \subref{k_smallworld} Small-world topology with $p=0.006$. The solid red line represents a fit to a $q$-Gaussian function with best-fitting obtained for $q= 1.65 \pm 0.01$ and $\beta = 2.211 \times 10^{-2} \pm 3.2 \times 10^{-4}$. The black solid line serves as guide to the eyes and has slope equal to $-2.80$. The number of vertices in this epicenters network is equal to 21\,157 and the numbers of edges 162\,992. \subref{k_regular} Regular topology. The solid red line represents a fit to a $q$-Gaussian function with best fitting obtained for $q=  2.61 \pm 0.01$ and $\beta = 7.673 \times 10^{-2} \pm 23.8 \times 10^{-4}$. The black solid line serves as guide to the eyes and has slope equal to $-1.22$. The number of vertices in this epicenters network equals 11\,642 and the number of edges 116\,560.}
    \label{k}
  \end{center}
\end{figure}

The result of this procedure is that, using the same conditions, the distribution obtained using small-world topologies  is different from that obtained when using regular topologies. As can be seen in Figure~\ref{k_smallworld}, where we used small-world topology, the cumulative distribution of connectivities follows a $q$-Gaussian with $q=1.65 \pm 0.01$, that can be approximated, by a power law $P(\geq k) \sim k^{-\delta}$ with $\delta = -2.80$, for connectivities $k>12$. 

For the regular case shown in Figure \ref{k_regular}, does not occur the same, since the distribution does not have a good agreement with the $q$-Gaussian function, and similarly, the power law approximation becomes inaccurate, illustrated by the fact that for the range where the $q$-Gaussian behaves as a power law there is a poor fit between the data and the fitted function. In~\cite{peixoto2006network} the authors used the regular topology and found an approximation a power-laws above a certain value of connectivity, but for this it is necessary much more larger lattices. In our study, only lattices sizes $L=200$ were necessary.
%
%podemos observar que para o intervalo onde a $q$-Gaussiana se comporta como uma lei de potência a aproximação não é válida tendo em vista que não há um bom ajustamento entre os dados e a lei de potência.
%
%tão pouco conseguimos realizar a aproximação de lei de potência para nenhum intervalo de conectividades, como pode ser visto em \ref{k_regular}.
%
It should also be noted that the $q$-Gaussian found in the small-world case agrees qualitatively with the result obtained in~\cite{ferreira2014towards} for a network of epicenters constructed from actual data. Another interesting similarity is that the $q$-Gaussian distribution also arises when analyzing the difference of magnitudes between successive earthquakes for both real data and synthetic data created from the OFC model with small-world topology, as reported in~\cite{caruso2007analysis}.

%Na Figure~\ref{k} temos as distribuições de conectividades para terremotos de tamanho $s \geq 50$. É importante destacar que a distribuição obtida na Figure~\ref{k_smallworld}, usando small-world lattice, é diferente da obtida na Figure~\ref{k_regular}, onde foi usado regular lattice. Para o caso small-world,  nós obtemos uma distribuição acumulada que obedece à uma $q$-Gaussiana com índice $q=1.65$.
%

\section{Conclusions}
In the present work we performed comparisons between results obtained through the use of catalogs produced by the OFC model for regular and small-world topologies. First, a study of the time interval between consecutive events was done thus obtaining the distributions of inter-event intervals. It was observed that the small-world case is in close agreement with results obtained using real data catalogs around the world, considering that both cases show distributions following $q$-exponentials. The same does not occur when we use regular topology, since the distribution obtained in this case does not have good fit to a $q$-exponential. After this, we studied the influence of the lattice topology on the border effect, that is, on how epicenters are distributed in lattice, where we observed that in the small-world, even for small probabilities of rewire ($p$), the intensity of the border effect is considerably lower than in the regular case. Furthermore, it was confirmed the influence of the probability $p$ in the measures of lattice's \textit{average path length} and \textit{diameter}, where can be observed that both characteristics have a fast drop for $p<0.1$, while, for $p>0.1$ remain approximately constant. Additionally we have built networks of successive epicenters and observed that in the small-world case the distribution of connectivities of vertices follows a $q$-Gaussian, and can be approximated by a power law for connectivities larger than 12. Again, we have an agreement with results obtained for real seismic catalogs across the globe because, similarly, the distribution of connectivities also obeys $q$-Gaussians which are approximated by power laws for higher values of connectivity. We highlight the fact that for the case of regular topology the fit to a $q$-Gaussian is not satisfactory, nor the power-law approximation. From our results it can be seen that the synthetic data generated using the OFC model with small-world topology has a better agreement with the results stemmed from actual data than the OFC model on a regular topology, which strengthens the idea that the Earth behaves as a self-organized critical system, and contribute to the conjecture of possible spatial and temporal long-range relationships in space and time between distantly located earthquakes.

%\green{Esse processo de deslizamento e relaxamento, passível de se transformar em uma reação em cadeia, será considerado como um terremoto.}

%\green{ Aqui nós iremos nos focar nos casos dissipativos, $0 < \alpha < 1/4$. }

%\green{Influência da topologia no modelo ( ajuda a se tornar SOC por exemplo, a partir do momento que passa a ter scaling, com relação ao parâmetro $\alpha$, na distribuição de magnitudes (no segundo artigo original do OFC tem scaling somente com relação ao tamanho, certo??) )...}

%\green{Influência das bordas. Maior incidência de eventos nas bordas. Citar: 1-OFC SOC scaling Paczuski2001 HIGHLIGHT.pdf  = Mostra que tem mais avalanche nas bordas. 2-OFC networks epicenters Prado2006.pdf  = Mostra que tem mais epicentros nas bordas. 3-OFC Distribution epicenters Prado2004.pdf  = Mostra que tem mais epicentros nas bordas.  }

%\green{Enquanto que para o caso regular  a distribuição acumulada de conectividades \red{possui um decaimento exponencial, $P_{reg}(\geq k) \sim e^{-\alpha k}$}, porém ao utilizar a small-world lattice observamos que ...}

\section{Acknowledgements}

A.R.R.P thanks CNPq (Brazilian Science Foundation) for his productivity fellowship.

\section{References}
\bibliographystyle{elsarticle-num} 
\bibliography{bib}


\end{document}